\documentstyle[12pt]{article}
\begin{document}

\title{\bf Interactions and Strings in  Higher Order Anisotropic
 and Inhomogeneous Superspaces and Isospaces\\
 ({\sf summary of monograph})}
\author{  \bf Sergiu  Vacaru}
\date{\sl [accepted by "Hadronic Press"]}
\maketitle
{\small
\centerline{\sl Academy of Sciences, Institute of Applied Physics,}
\centerline{{\sl Chi\c sin\v au 2028,} {\bf Republic of Moldova}}
\centerline{ and }
\centerline {\sl Institute for Basic Research,}
\centerline{{\sf Palm Harbor,} {\bf USA}\ \& \ {\sf Molise,} {\bf Italy}}
\vskip3pt
\centerline{ e-mail: lises@cc.acad.md}}
\vskip20pt
\centerline{\bf Preface}

This monograph gives a general geometric background of
the theory of field interactions, strings and diffusion processes on
spaces, superspaces and isospaces with higher order anisotro\-py and
 inhomogenity.
 The last fifteen years have been attempted
 a number of extensions of Finsler geometry with applications
 in theoretical and mathematical physics and biology which go far beyond the
 original scope. Our approach proceeds by  developing the concept of higher
 order anisotropic superspace which unify the logical and mathematical aspects
 of modern Kaluza--Klein theories and generalized Lagrange and Finsler
 geometry and leads to
 modelling of physical processes on higher order fiber bundles provided
 with nonlinear and distingushed connections and metric structures. The view
adopted here is that a general field theory should incorporate
 all possible anisotropic and stochastic manifestations of classical and
 quantum interactions and, in consequence, a corresponding modification
 of basic principles and mathematical methods in formulation of physical
 theories. This monograph can be also considered as a pedagogical survey on
 the mentioned subjects.

 There are established
three approaches for modeling  field interactions and spaces
anisotropies. The first one is to deal with a usual locally isotropic
physical theory and to consider anisotropies as a consequence of the
anisotropic structure of sources in field equations (for instance, a number
of cosmological models are proposed in the framework of the Einstein theory
with the energy--momentum generated by anisotropic matter, as a general
reference see 
 [165]). The second approach to
anisotropies originates from the Finsler geometry
 [78,55,213,159] and its generalizations
 [17,18,19,160,161,13,29, 256, 255, 264, 272]
 with a general
imbedding into Kaluza--Klein (super) gravity and string theories
 [269,270,260,265,266,267],
and speculates a generic
anisotropy of the space--time structure and of fundamental field of
interactions. The Santilli's approach
 [217,219,220,218,221,\\ 222,223,224] is more radical by proposing a
generalization of Lie theory and introducing isofields, isodualities and
related mathematical structures. Roughly speaking, by using corresponding
partitions of the unit we can model possible metric anisotropies as in
Finsler or generalized Lagrange geometry but the problem is also to take
into account classes of anisotropies generated by nonlinear and
distinguished connections.

In its present version this book addresses itself both to mathematicians and
physicists, to researches and graduate students which are interested in
geometrical aspects of fields theories, extended (super)gravity and
(super)strings and supersymmetric diffusion. It presupposes a general
background in the mentioned divisions of modern theoretical physics and
 assumes some familiarity with differential geometry, group theory, complex
 analysis and stochastic calculus.

The monograph is divided into three parts:

 The first five Chapters cover
the  higher order anisotropic supersymmetric theories:  Chapter 1
 is devoted to the geometry of higher order anisotropic supersaces with
an extension to supergravity models in Chapter 2. The supersymmetric nearly
autoparallel maps of superbundles and higher order anisotropic conservation
 laws are considered in Chapter 3. Higher order anisotropic superstrings
and anomalies are studied in Chapter 4. Chapter 5 contains an introduction
 into the theory of supersymmetric locally anisotropic stochastic processes.

 The next five Chapters are devoted to the (non supersymmetric) theory
of higher order anisotropic interactions and  stochastic processes. Chapter 6
 concerns the spinor formulation of field theories with locally
 anisotropic interactions and Chapter 7 considers anisotropic gauge field and
 gauge gravity models. Defining nearly autoparallel maps as generalizations
 of conformal transforms we analyze the problem of formulation of conservation
 laws in higher order anisotropic spaces in Chapter 8. Nonlinear sigma models
 and strings in locally anisotropic backgrounds are studied in Chapter 9.
 Chapter 10 is devoted to the theory of stochastic differential equations for
 locally anisotropic diffusion processes.

The rest four Chapters presents a study on Santilli's locally anisotropic
 and inhomogeneous isogeometries, namely, an introduction into the
 theory of isobuncles and generalized isofinsler gravity.
Chapter 11 is devoted to basic notations and definitions on Santilli and
coauthors isotheory. We introduce the bundle isospaces in Chapter 12 where
some necessary properties of Lie--Santilli isoalgebras and isogroups and
corresponding isotopic extensions of manifolds are applied in order to
define fiber isospaces and consider their such (being very important for
modeling of isofield interactions) classes of principal isobundles and
vector isobundles. In that Chapter there are also studied the
 isogeometry of nonlinear isoconnections in vector
 isobundles, the isotopic distinguishing of geometric
objects, the isocurvatures and isotorsions of nonlinear and distinguished
 isoconnections and the
structure equations and invariant conditions.  The
next Chapter 13 is devoted to the isotopic extensions of generalized Lagrange
and Finsler geometries. In Chapter 14 the isofield equations of locally
 anisotropic and inhomogeneous interactions will be analyzed and  an outlook
 and conclusions will be presented.

We have not attempted to give many details on previous knowledge of the
subjects or complete list of references. Each Chapter contains a brief
introduction, the first section  reviews the basic results, original papers
and monographs. If it is considered necessary, outlook and discussion are
presented at the end of the Chapter.

We hope that the reader will not suffer too much from our insufficient
 mastery of the English language.
\vskip15pt

{\bf Acknowledgments.}

 It is a pleasure for the author to give many thanks
especially to Professors R. Miron, M.Anastasiei, R. M. Santilli and
 A. Bejancu  for valuable discussions, collaboration
 and necessary offprints.

 The warmest thanks are extended to Drs
  E.  Seleznev  and L. Konopko for their help and support.

The author wish to express generic thanks to the referees for a detailed
control and numerous constructive suggestions as well he
should like to express his deep gratitude to the publishers for their
unfailing support.
\vskip20pt
\centerline{Part I. {\bf Higher Or\-der Anisotropic Supersymmetry}}
\vskip10pt
\centerline{Capter 1. {\bf HA--Superspaces}}
\vskip5pt
The differential supergeometry have been formulated with the aim of getting
a geometric framework for the supersymmetric field theories (see the theory
of graded manifolds  [37,146,147,144], the theory of supermanifolds
 [290,203,27,127] and, for detailed considerations of geometric and
topological aspects of supermanifolds and formulation of superanalysis,
 [63,49,157,\\ 114,281,283]). In this Chapter we apply the supergeometric
formalism for a study of a new class of (higher order anisotropic)
superspaces.

The concept of local anisotropy is largely used in some divisions of
theoretical and mathematical physics  [282,119,122,163] (see also
 possible applications in physics and biology in
 [14,13]). The first
models of locally anisotropic (la) spaces (la--spaces) have been proposed by
P.Finsler
 [78] and E.Cartan
 [55] (early approaches and modern
treatments of Finsler geometry and its extensions can be found, for
instance, in
 [213,17,18,159]). In our works
 [256,255,258,259,260,264,272,279,276] we try to formulate the
geometry of la--spaces in a manner as to include both variants of Finsler and
Lagrange, in general supersymmetric, extensions and higher dimensional
Kaluza--Klein (super)spaces as well to propose general principles and
methods of construction of models of classical and quantum field interactions
and stochastic processes on spaces with generic anisotropy.

We cite here the works
 [31,33] by A. Bejancu where a new
viewpoint on differential geometry of supermanifolds is considered. The
author introduced the nonlinear connection (N--connection) structure and
developed a corresponding distinguished by N--connection supertensor
covariant differential calculus in the frame of De Witt 
 [290] approach
to supermanifolds in the framework of the geometry of superbundles with
typical fibres parametrized by noncommutative coordinates. This was the
first example of superspace with local anisotropy. In our turn we have given
a general definition of locally anisotropic superspaces (la--superspaces)
 [260]. It should be noted here that in our supersymmetric
generalizations  we were inspired by the R. Miron, M. Anastasiei and Gh.
Atanasiu works on the geometry of nonlinear connections in vector bundles
and higher order La\-gran\-ge spaces 
 [160,161,162].
In this Chapter we shall formulate the theory of higher order vector and
tangent superbundles provided with nonlinear and distinguished connections
and metric structures (a generalized model of la--superspaces). Such
superbundles contain as particular cases the supersymmetric extensions and
various higher order prolongations of Riemann, Finsler and Lagrange spaces.
We shall use instead the terms distinguished superbundles, distinguished
 geometric objects and so on (geometrical constructions distinguished
 by a N--connection structure) theirs corresponding brief denotations
 (d--superbundles,  d--objects and so on).

The Chapter is organized as follows: Section 1.1 contains a brief review on
supermanifolds and superbundles and an introduction into the geometry of
higher order distinguished vector superbundles. Section 1.2 deals with the
geometry of nonlinear and linear distinguished connections in vector
superbundles and distinguished vector superbundles. The geometry of the
total space of distinguished vector superbundles is studied in section 1.3;
distinguished connection and metric structures,  their torsions, curvatures
and structure equations are considered. Generalized Lagrange and Finsler
superspaces there higher order prolongations are  defined in section 1.4 .
Concluding remarks on Chapter 1 are contained in section 1.5.
\vskip10pt
\centerline{Chapter 2. {\bf HA--Supergravity}}
\vskip5pt
In this Chapter we shall analyze three models of supergravity with higher
order anisotropy. We shall begin our considerations with N--connection
s--spaces in section 2.1. Such s--spaces are generalizations of flat
s--spaces containing a nontrivial N--connection structure but with vanishing
d--con\-nec\-ti\-on. We shall introduce locally
adapted s--vielbeins and define
s--fields and differential forms in N--connection s--spaces. Sections 2.2 and
2.3 are correspondingly devoted to gauge s-field and s--gravity theory in
osculator s--bundles. In order to have the possibility to compare our model
with usual {\sf N}=1 ( one dimensional supersymmetric extensions; see, for
instance, 
 286,170,215]) supergravitational models we develop a
supergavity theory on osculator s--bundle $Osc^z\widetilde{M}_{(M)}$ where
the even part of s--manifold $\widetilde{M}_{(M)}$ has a local structure of
Minkowski space with action of Poincare group. In this case we do not have
problems connected with definition of spinors (Lorentz, Weyl or Maiorana
type) for spaces of arbitrary dimensions and can solve Bianchi identities.
As a matter of principle, by using our results on higher dimensional and
locally anisotropic spaces, see 
 [256,255] we can introduce
distinguished spinor stuctures and develop variants of extended supergravity
with general higher order anisotropy. This approach is based on global
geometric constructions and allows us to avoid tedious variational
calculations and define the basic field equations and conservation laws on
s--spaces with local anisotropy. That why, in section 2.5, we introduce
Einstein--Cartan equations on distinguished vector superbundles (locally
parametrized by arbitrary both type commuting and anticommuting coordinates)
in a geometric manner, in some line following the geometric background for
Einstein realtivity, but in our case on dvs-bundels provided with arbitrary
N--connection and distinguished torsion and metric structures. We can
consider different models, for instance, with prescribed N--connection and
torsions, to develop a Einstein--Cartan like theory, or to follow approaches
from gauge gravity. In section 2.6 we propose a variant of gauge like higher
anisotropic supergravity being a generalization to dvs--bundles of models of
locally anisotropic gauge gravity  [258,259,272] (see also Chapter 7
in this monograph).
\vskip10pt
\centerline{Chapter 3. {\bf Supersymmetric NA--Maps}}
\vskip10pt
The study of models of classical and quantum field interactions in higher
dimension superspaces with, or not, local anisotropy is in order of the day.
The development of this direction entails great difficulties because of
problematical character of the possibility and manner of definition of
conservation laws on la--spaces. It will be recalled that conservation laws
of energy--momentum type are a consequence of existence of a global group of
automorphisms of the fundamental Mikowski spaces. As a rule one considers
the tangent space's automorphisms or symmetries conditioned by the existence of
Killing vectors on curved (pseudo)Riemannian spaces. There are not any
global or local automorphisms on generic la--spaces and in result of this
fact, at first glance, there are a lot of substantial difficulties with
formulation of conservation laws and, in general, of physical consistent
field theories with local anisotropy. R. Miron and M. Anastasiei
investigated the nonzero divergence of the matter energy--momentum d--tensor,
the source in Einstein equations on la--spaces, and considered an original
approach to the geometry of time--dependent Lagrangians
 [12,160,161]. In a series of papers
 [249,276,263,273,274,278,279,252,275,277] we attempt to solve
the problem of definition of energy-momentum values for locally isotropic
and anisotropic gravitational and matter fields interactions and of
conservation laws for basic physical values on spaces with local anisotropy
in the framework of the theory of nearly geodesic and nearly autoparallel
maps.

In this Chapter a necessary geometric background (the theory of nearly
autoparallel maps, in brief na-maps, and tensor integral formalism) for
formulation and investigation of conservation laws on higher order isotropic
and anisotropic superspaces is developed. The class of na--maps contains as a
particular case the conformal transforms and is characterized by corresponding
 invariant conditions for generalized Weyl tensors and Thomas parameters
  [227,230]. We can connect the na--map theory with the formalism of
tensor integral and
multitensors on distinguished vector superbundles. This approaches based
 on generalized conformal transforms of superspaces with
 or not different types of higher order anisotropy consist a new division of
 differential supergeometry with applications in modern theoretical and
 mathematical physics.

We note that in most cases proofs
of our theorems are mechanical but rather tedious calculations similar to
those presented in  [230,252,263]. Some of them will be given in
detail, the rest will be sketched. We shall omit splitting of formulas into
even and odd components (see Chapter 8 on nearly autoparallel maps and
conservation laws for higher order (non supersymmetric) anisotropic spaces).

Section 3.1 is devoted to the formulation of the theory of nearly
autoparallel maps of dvs--bundles. The classification of na--maps and
formulation of their invariant conditions are given in section 3.2. In
section 3.3 we define the nearly autoparallel tensor--integral on locally
anisotropic multispaces. The problem of formulation of conservation laws on
spaces with local anisotropy is studied in section 3.4. Some conclusions are
 presented in section 3.5.
\vskip10pt
\centerline{Chapter 4. {\bf HA--Superstrings}}
\vskip5pt
The superstring theory holds the greatest promise as the unification theory
of all fundamental interactions. The superstring models contains a lot a
characteristic features of Kaluza--Klein approaches, supersymmetry and
supergravity, local field theory and dual models. We note that in the string
theories the nonlocal one dimensional quantum objects (strings) mutually
interacting by linking and separating together are considered as fundamental
values. Perturbations of the quantized string are identified with quantum
particles. Symmetry and conservation laws in the string and superstring
theory can be considered as sweeping generalizations of gauge principles
which consists the basis of quantum field models. The new physical concepts
are formulated in the framework a ''new'' for physicists mathematical
formalism of the algebraic geometry and topology 
 [106].

The relationship between two dimensional $\sigma $-models and strings has
been considered
 [153,80,53,229,7] in order to discuss the
effective low energy field equations for the massless models of strings.
Nonlinear $\sigma $-models makes up a class of quantum field systems for
which the fields are also treated as coordinates of some manifolds.
Interactions are introduced in a geometric manner and admit a lot of
applications and generalizations in classical and quantum field and string
theories. The geometric structure of nonlinear sigma models manifests the
existence of topological nontrivial configuration, admits a geometric
interpretation of conterterms and points to a substantial interrelation
between extended supersymmetry and differential supergeometry. In connection
to this a new approach based on nonlocal, in general, higher order
anisotropic constructions seem to be emerging 
 [254,269,261]. We
consider the reader to be familiar with basic results from supergeometry
(see, for instance,
 [63,147,290,203]), supergravity theories
 [86,215,288,286,287] and superstrings 
 [115,289,139,140].

In this Chapter we shall present an introduction into the theory of higher
order anisotropic superstrings being a natural generalization to locally
anisotropic (la) backgrounds (we shall write in brief la-backgrounds,
la-spaces and la-geometry) of the Polyakov's covariant functional-integral
approach to string theory 
 [193]. Our aim is to show that a
corresponding low-energy string dynamics contains the motion equations for
field equations on higher order anisotropic superspaces and to analyze the
geometry of the perturbation theory of the locally anisotropic
supersymmetric sigma models. We note that this Chapter is devoted to
supersymmetric models of locally anisotropic superstrings (details on the
so called bosonic higher order anisotropic strings are given in the Chapter
9).

The plan of presentation in the Chapter is as follows. Section 4.1 contains
an introduction into the geometry of two dimensional higher order
anisotropic sigma models and an locally anisotropic approach to heterotic
strings. In section 4.2 the background field method for $\sigma $-models is
generalized for a distinguished calculus locally adapted to the
N--connection structure in higher order anisotropic superspaces. Section 4.3
is devoted to a study of Green--Schwartz action in distinguished vector
superbundles. Fermi strings in higher order anisotropic spaces are considered
in section 4.4. An example of one--loop and two--loop calculus for anomalies
of locally anisotropic strings is presented in section 4.5. Conclusions are
drawn in section 4.6.\
\vskip20pt
\centerline{Chapter 5. {\bf Stochastics in LAS--Spaces}}
\vskip10pt
We shall describe the analytic results which combine the fermionic Brownian
motion with stochastic integration in higher order anisotropic spaces. It
will be shown that a wide class of stochastic differential equations in
locally anisotropic superspaces have solutions. Such solutions will be than
used to derive a Feynman--Kac formula for higher order anisotropic systems.
We shall achieve this by introducing locally anisotropic superpaths
parametrized by a commuting and an anticommuting time variable. The
supersymmetric stochastic techniques employed in this Chapter was developed
by A. Rogers in a series of works 
 [206,207,205,209,210]
(superpaths have been also considered in papers
 [103,82]
 and 
 [198]). One of
the main our purposes is to extend this formalism in order to formulate the
theory of higher order anisotropic processes in distinguished vector
superbundles 
 [260,262,265,266,267,253,268].
Stochastic calculus for bosonic and fermionic Brownian paths will provide a
geometric approach to Brownian motion in locally anisotropic superspaces.

Sections 5.1 and 5.2 of this Chapter contain correspondingly a brief
introduction  into the subject and a brief review of fermionic Brownian
motion and path  integration. Section 5.3 considers distinguished stochastic
integrals in the  presence of fermionic paths. Some results in calculus on a
(1,1)--dimensional superspace and two supersymmetric formulae for superpaths
are described in section 5.4 and than, in section 5.5,  the theorem on the
existence of unique  solutions to a useful class of distinguished
stochastic differential  equations is proved and the distinguished
supersymmetric Feynman--Kac formula  is established. Section 5.6 defines
some higher order anisotropic manifolds  which can be constructed from a
vector bundle over a vector  bundle provided with compatible nonlinear and
distinguished connection and metric structures. In section 5.7 a geometric
formulation of Brownian paths on  higher order anisotropic manifolds is
contained; these paths are used to give  a Feynman--Kac formula for the
Laplace--Beltrami operator for twisted  differential forms. This formula is
used to give a proof of the index theorem  using supersymmetry of the higher
order anisotropic superspaces in section 5.8  (we shall apply the methods
developed in
 [4,82]
  and 
 [209,210].
\vskip20pt
\centerline{Part II. {\bf Higher Order Anisotropic Interac\-ti\-ons}}
\vskip10pt
\centerline{Chapter 6. {\bf HA--Spinors}}
\vskip5pt
Some of fundamental problems in physics advocate the extension to locally
anisotropic and higher order anisotropic backgrounds of physical theories
 [159,161,13,29,18,162,272,265,266,267]. In order to
construct physical models on higher order anisotropic spaces it is necessary
a corresponding generalization of the spinor theory. Spinor variables and
interactions of spinor fields on Finsler spaces were used in a heuristic
manner, for instance, in works
 [18,177], where the problem of a
rigorous definition of la-spinors for la-spaces was not considered. Here we
note that, in general, the nontrivial nonlinear connection and torsion
structures and possible incompatibility of metric and connections makes the
solution of the mentioned problem very sophisticate. The geometric
definition of la-spinors and a detailed study of the relationship between
Clifford, spinor and nonlinear and distinguished connections structures in
vector bundles, generalized Lagrange and Finsler spaces are presented in
refs. 
 [256,255,264].

The purpose of the Chapter is to summarize and extend our investigations
 [256,255,264,272,260] on formulation of the theory of classical
and quantum field interactions on higher order anisotropic spaces. We
receive primary attention to the development of the necessary geometric
framework: to propose an abstract spinor formalism and formulate the
differential geometry of higher order anisotropic spaces . The next step is
the investigation of higher order anisotropic interactions of fundamental
fields on generic higher order anisotropic spaces (in brief we shall use
instead of higher order anisotropic the abbreviation ha-, for instance,
ha--spaces, ha--interactions and ha--spinors).

In order to develop the higher order anisotropic spinor theory it will be
convenient to extend the Penrose and Rindler abstract index formalism
 [180,181,182] (see also the Luehr and Rosenbaum index free methods
 [154]) proposed for spinors on locally isotropic spaces. We note that in
order to formulate the locally anisotropic physics usually we have
dimensions $d>4$ for the fundamental, in general higher order anisotropic
space--time and to take into account physical effects of the nonlinear
connection structure. In this case the 2-spinor calculus does not play a
 preferential role.

Section 6.1 of this Chapter contains an introduction into the geometry of higher
order anisotropic spaces, the  distinguishing of geometric objects by
N--connection structures  in such spaces is analyzed, explicit formulas for
coefficients of torsions and curvatures of N- and d--connections are presented
and the field equations for gravitational interactions with higher order
anisotropy are formulated. The distinguished Clifford algebras are introduced
in section 6.2 and higher order anisotropic Clifford bundles are defined in
 section 6.3. We present a study of almost complex structure for the case of
 locally anisotropic spaces modeled in the framework of the almost Hermitian
 model of generalized Lagrange spaces in section 6.4. The d--spinor
 techniques is analyzed in section 6.5 and the differential
 geometry of higher order anisotropic spinors is formulated in section 6.6.
 The section 6.7 is devoted to geometric aspects of the theory of field
 interactions with higher order anisotropy (the d--tensor and d--spinor form
 of basic field equations for gravitational, gauge and d--spinor fields are
 introduced). Finally, an outlook and conclusions on ha--spinors are given in
 section 6.8.
\vskip20pt
\centerline{Chapter 7. {\bf Gauge and Gravitational Ha--Fields}}
\vskip10pt
Despite the charm and success of general relativity there are some
fundamental problems still unsolved in the framework of this theory. Here we
point out the undetermined status of singularities, the problem of
formulation of conservation laws in curved spaces, and the
unrenormalizability of quantum field interactions. To overcome these defects
a number of authors (see, for example, Refs.  [240,285,194,3])
tended
to reconsider and reformulate gravitational theories as a gauge model
similar to the theories of weak, electromagnetic, and strong forces. But, in
spite of theoretical arguments and the attractive appearance of different
proposed models of gauge gravity, the possibility and manner of
interpretation of gravity as a kind of gauge interaction remain unclear.

The work of Popov and Daikhin 
 [195,196] is distinguished among other gauge
approaches to gravity. Popov and Dikhin did not advance a gauge extension,
or modification, of general relativity; they obtained an equivalent
reformulation (such as well-known tetrad or spinor variants) of the Einstein
equations as Yang-Mills equations for correspondingly induced Cartan
connections
 [40] in the affine frame bundle on the pseudo-Riemannian
space time. This result was used in solving some specific problems in
mathematical physics, for example, for formulation of a twistor-gauge
interpretation of gravity and of nearly autoparallel conservation laws on
curved spaces 
 [246,250,252,249,233]. It has also an important
conceptual role. On one hand, it points to a possible unified treatment of
gauge and gravitational fields in the language of linear connections in
corresponding vector bundles. On the other, it emphasize that the types of
fundamental interactions mentioned essentially differ one from another, even
if we admit for both of them a common gauge like formalism, because if to
Yang-Mills fields one associates semisimple gauge groups, to gauge
treatments of Einstein gravitational fields one has to introduce into
consideration nonsemisimple gauge groups.

Recent developments in theoretical physics suggest the idea that a more
adequate description of radiational, statistical, and relativistic optic
effects in classical and quantum gravity requires extensions of the
geometric background of theories
 [282,163,13,14,15,28,29,118,119,120,122,212,235,236,238,\\ 280,41] by
introducing into consideration spaces with local anisotropy and formulating
corresponding variants of Lagrange and Finsler gravity and theirs extensions
 to higher order anisotropic spaces 
 [295,162,266,267,268].

The aim of this Chapter is twofold. The first objective is to present our
results 
 [272,258,259] on formulation of geometrical approach to
interactions of Yang-Mills fields on spaces with higher order anisotropy in the
framework of the theory of linear connections in vector bundles (with
semisimple structural groups) on ha-spaces. The second objective is to
extend the geometrical formalism in a manner including theories with
nonsemisimple groups which permit a unique fiber bundle treatment for both
locally anisotropic Yang-Mills field and gravitational interactions. In
general lines, we shall follow the ideas and geometric methods proposed in
refs.
 [240,195,196,194,40] but we shall apply them in a form convenient
for introducing into consideration geometrical constructions
 [160,161] and physical theories on ha-spaces.

There is a number of works on gauge models of interactions on Finsler spaces
and theirs extensions(see, for instance,
 [17,18,19,28,164,177]). One has introduced different variants of
generalized gauge transforms, postulated corresponding Lagrangians for
gravitational, gauge and matter field interactions and formulated
variational calculus (here we note the approach developed by A. Bejancu
 [30,32,29]). The main problem of such models is the dependence of
the basic equations on chosen definition of gauge "compensation" symmetries
and on type of space and field interactions anisotropy. In order to avoid
the ambiguities connected with particular characteristics of possible
la-gauge theories we consider a "pure" geometric approach to gauge theories
(on both locally isotropic and anisotropic spaces) in the framework of the
theory of fiber bundles provided in general with different types of
nonlinear and linear multiconnection and metric structures. This way, based
on global geometric methods, holds also good for nonvariational, in the
total spaces of bundles, gauge theories (in the case of gauge gravity based
on Poincare or affine gauge groups); physical values and motion (field)
equations have adequate geometric interpretation and do not depend on the
type of local anisotropy of space-time background. It should be emphasized
here that extensions for higher order anisotropic spaces which will be
presented in this Chapter can be realized in a straightforward manner.

The presentation in the Chapter is organized as follows:

In section 7.1 we give a geometrical interpretation of gauge (Yang-Mills)
fields on general ha-spaces. Section 7.2 contains a geometrical definition
of anisotropic Yang-Mills equations; the variational proof of gauge field
equations is considered in connection with the "pure" geometrical method of
introducing field equations. In section 7.3 the ha--gravity is reformulated
as a gauge theory for nonsemisimple groups. A model of nonlinear de Sitter
gauge gravity with local anisotropy is formulated in section 7.4. We study
gravitational gauge instantons with trivial local anisotropy in section 7.5.
Some remarks are given in section 7.6.
\vskip20pt
\centerline{Chapter 8. {\bf Na--Maps and Conservation Laws}}
\vskip10pt
Theories of field interactions on locally anisotropic curved spaces form a
new branch of modern theoretical and mathematical physics. They are used for
modelling in a self--consistent manner physical processes in locally
anisotropic, stochastic and turbulent media with beak radiational reaction
and diffusion
 [161,13,14,282]. The first model of locally
anisotropic space was proposed by P.Finsler 
 [78] as a generalization
of  Riemannian geometry; here we also cite the fundamental contribution made
by E. Cartan [55] and mention that in monographs [213,159,17,19,29]
detailed  bibliographies are contained. In this
Chapter we follow R. Miron and M. Anastasiei
 [160,161] conventions and
base our investigations on their general model of locally anisotropic (la)
gravity (in brief we shall write la-gravity) on vector bundles, v--bundles,
provided with nonlinear and distinguished connection and metric structures
(we call a such type of v--bundle as a la-space if connections and metric
are compatible).

The study of models of classical and quantum field interactions on la-spaces
is in order of the day. For instance, the problem of definition of spinors
on la-spaces is already solved (see 
 [256,275,264] and Chapter
6 and some models of locally anisotropic Yang--Mills and gauge like
gravitational interactions are analyzed (see 
 [272,263] and Chapter 7
and alternative approaches in
 [17,29,122,119]). The development of
this direction entails great difficulties because of problematical character
of the possibility and manner of definition of conservation laws on
la-spaces. It will be recalled that, for instance, conservation laws of
energy--momentum type are a consequence of existence of a global group of
automorphisms of the fundamental Mikowski spaces (for (pseudo)Riemannian
spaces the tangent space' automorphisms and particular cases when there are
symmetries generated by existence of Killing vectors are considered). No
global or local automorphisms exist on generic la-spaces and in result of
this fact the formulation of la-conservation laws is sophisticate and full
of ambiguities. R. Miron and M. Anastasiei firstly pointed out the nonzero
divergence of the matter energy-momentum d--tensor, the source in Einstein
equations on la-spaces, and considered an original approach to the geometry
of time--dependent Lagrangians
 [12,160,161]. Nevertheless, the
rigorous definition of energy-momentum values for la-gravitational and
matter fields and the form of conservation laws for such values have not
been considered in present--day studies of the mentioned problem.

The aim of this Chapter is to develop a necessary geometric background (the
theory of nearly autoparallel maps, in brief na-maps,  and tensor integral
formalism on la-multispaces) for formulation and a detailed investigation of
conservation laws on locally isotropic and anisotropic curved spaces. We
shall summarize our results on formulation of na-maps for generalized affine
spaces (GAM-spaces) 
 [249,251,273], Einstein-Cartan and Einstein spaces
  [250,247,278] bundle spaces 
  [250,247,278] and different
classes of la-spaces [279,276,102,263]
 and present an extension of
the na-map theory for superspaces. For simplicity we shall restrict our
considerations only with the "first" order anisotropy (the basic  results on
higher order anisotropies a presented in a supersymmetric manner  Chapter
3. Comparing the geometric constructions from both Chapters on  na--map
theory we assure ourselves that the developed methods hold good  for all
type of curved spaces (with or not torsion, locally isotropic or  even with
local anisotropy and being, or not, supesymmetric). In order  to make the
reader more familiar with na--maps and theirs applications  and to point to
some common features, as well to proper particularities of  the
supersymmetric and higher order constructions, we shall recirculate some
basic definitions, theorems and proofs from Chapter 3.

The question of definition of tensor integration as the inverse operation of
covariant derivation was posed and studied by A.Mo\'or 
 [167].
Tensor--integral and bitensor formalisms turned out to be very useful in
solving certain problems connected with conservation laws in general
relativity
 [100,247].  In order to extend tensor--integral
constructions we have proposed 
 [273,278] to  take into consideration
nearly autoparallel
 [249,247,250] and nearly geodesic 
 [230] maps, ng--maps,
 which forms a subclass of local 1--1 maps of  curved spaces
with deformation of the connection and metric structures. A generalization
of the Sinyukov's ng--theory for spaces with local anisotropy was proposed
by considering maps with deformation of connection for Lagrange spaces (on
Lagrange spaces see
 [136,160,161]) and generalized Lagrange spaces
 [263,279,276,275,101]. Tensor integration formalism for  generalized
Lagrange spaces was developed in
[255,102,263]. One of the main
purposes of this Chapter is to synthesize the results obtained in the
mentioned  works and to formulate them for a very general class of
la--spaces. As the next step the la--gravity and analysis of
la--conservation laws are considered.

We note that proofs of our theorems are mechanical, but, in most cases, they
are rather tedious calculations similar to those presented in
 [230,252,263]. Some of them, on la-spaces, will be given in detail the
rest, being similar, or consequences, will be only sketched or omitted.

Section 8.1 is  devoted to the formulation of the theory of nearly
autoparallel maps of  la--spaces. The classification of na--maps and
formulation of their invariant conditions are given in section 8.2. In
section 8.3 we define the nearly  autoparallel tensor--integral on locally
anisotropic multispaces. The problem of formulation of conservation laws on
spaces with local anisotropy is studied in section 8.4 . We present a
definition of conservation laws for la--gravitational fields on na--images
of la--spaces in section 8.5. Finally,  in this Chapter, section 8.6, we
analyze the locally isotropic limit, to  the Einstein gravity and it
generalizations, of the na-conservation laws.
\vskip20pt
\centerline{Chapter 9. {\bf HA--Strings}}
\vskip10pt
The relationship between two dimensional $\sigma $-models and strings has
been considered
 [153,80,53,229,7]
in order to discuss the
effective low energy field equations for the massless models of strings. In
this Chapter we shall study some of the problems associated with the theory
of higher order anisotropic strings being a natural generalization to higher
order anisotropic  backgrounds (we shall write in brief ha--backgrounds,
ha--spaces and ha--geometry) of the Polyakov's covariant functional--integral
approach to string theory 
 [193]. Our aim is to show that a
corresponding low--energy string dynamics contains the motion equations for
field equations on ha-spaces; models of ha--gravity could be more adequate
for investigation of quantum gravitational and Early Universe cosmology.

The plan of the Chapter is as follows. We begin, section 9.1, with a study
of the nonlinear $\sigma$--model and ha--string propagation by developing the
d--covariant method of ha--background field. Section 9.2 is devoted to
problems of regularization and renormalization of the locally anisotropic $%
\sigma$--model and a corresponding analysis of one- and two--loop diagrams of
this model. Scattering of ha-gravitons and duality are considered in section
9.3, and a summary and conclusions are drawn in section 9.4.\
\vskip20pt
\centerline{Chapter 10. {\bf Stochastic Processes on HA--Spaces}}
\vskip10pt
The purpose of investigations in this Chapter 
 [253,262] is to
extend the formalism of stochastic calculus to the case of spaces with
\index{Stochastic Processes!on ha--spaces}
higher order anisotropy (distinguished vector bundles with compatible
nonlinear and distinguished connections and metric structures and
generalized Lagrange and Finsler spaces). We shall examine nondegenerate
diffusions on the mentioned spaces and theirs horizontal lifts.

Probability theorists, physicists, biologists and financiers are already
familiar with classical and quantum statistical and geometric methods
applied in va\-ri\-ous bran\-ches of science and economy
 [13,14,171,189,131,84,141]. We note that modeling of diffusion processes
in nonhomogerneous media and formulation of nonlinear thermodynamics in
physics, or of dynamics of evolution of species in biology, requires a more
extended geometrical background than that used in the theory of stochastic
differential equations and diffusion processes on Riemann, Lorentz
manifolds 
 [117,74,75,76] and in Rieman--Cartan--Weyl spaces 
 [197,199].

Our aim is to formulate the theory of diffusion processes on spaces with
local anisotropy. As a model of such spaces we choose vector bundles on
space-times provided with nonlinear and distinguished connections and metric
structures
 [160,161]. Transferring our considerations on tangent
bundles we shall formulate the theory of stochastic differential equations
on generalized Lagrange spaces which contain as particular cases Lagrange
and Finsler spaces
 [213,17,18,29,159].

The plan of the presentation in the Chapter is as follow: We present a brief
introduction into the theory of stochastic differential equations and
diffusion processes on Euclidean spaces in section 10.1. In section 10.2 we
give a brief summary of the geometry of higher order anisotropic spaces.
Section 10.3 is dedicated to the formulation of the theory of stochastic
differential equations in distinguished vector bundle spaces. This section
also concerns the basic aspects of stochastic calculus in curved spaces. In
section 10.4 the heat equations in bundle spaces are analyzed. The
nondegenerate diffusion on spaces with higher order anisotropy is defined in
section 10.5. We shall generalize in section 10.6 the results of section
10.4 to the case of heat equations for distinguished tensor fields in vector
bundles with (or not) boundary conditions. Section 10.7 contains concluding
remarks and a discussion of the obtained results.\
\vskip20pt
\centerline{Part III. {\bf Isobundles and Generalized
Isofinsler Gravity }}
\vskip10pt
\centerline{Chapter 11. {\bf Basic Notions on Isotopies}}
\vskip5pt
 This Part is devoted to a generalization 
 [271] of the geometry of
 Santilli's locally anisotropic and inhomogeneous
 isospaces 
  [217,219,220,218,221,222,\\ 223,224]
 to the geometry of vector isobundles  provided with nonlinear and
 distinguished isoconnections and isometric  structures. We present,
 apparently for the first time, the isotopies of Lagrange,
Finsler and Kaluza--Klein
 spaces. We also continue the study of the
 interior, locally anisotropic and inhomogeneous gravitation by extending
 the isoriemannian space's constructions and presenting a geometric
 background  for the theory of isofield interactions  in
  generalized isolagrange and isofinsler spaces.

 The main purpose of this Part is to formulate a synthesis of
the Santilli isotheory and the approach on modeling locally anisotropic
geometries and physical models on bundle spaces provided with nonlinear
connection and distinguished connection and metric structures
 [160,161,295].
 The isotopic
variants of generalized Lagrange and Finsler geometry will be analyzed.
Basic geometric constructions such as nonlinear isoconnections in vector
isobundles, the isotopic curvatures and torsions of distinguished
isoconnections
and theirs structure equations and invariant values will be defined. A model
of locally anisotropic and inhomogeneous
gravitational isotheory will be constructed.

 Our study of Santilli's isospaces and isogeometries over isofields  will
 be treated via the isodifferential calculus according to their
latest formulation 
 [224] (we extend this calculus
 for isospaces provided with nonlinear isoconnection structure). We shall
also use Kadeisvili's notion of isocontinuity
 [129,130] and the
novel Santilli--Tsagas--Sourlas isodifferential topology
 [223,239,232].

After reviewing the basic elements for completeness as well as for notational
convenience, we shall extend  Santilli's foundations of the isosympletic
 geometry 
 [223] to isobundles and related aspects
 (by applying, in an isotopic manner, the methods summarized in
 Miron and Anastasiei  
 [160,161] and
 Yano and Ishihara 
 [295] monographs).
 We shall  apply our results
on isotopies of Lagrange, Finsler and Kaluza--Klein geometries
 to further
 studies of the isogravitational theories (for isoriemannian spaces
 firstly considered by Santilli 
 [223]) on vector isobundle
 provided with
 compatible nonlinear and distinguished isoconnections and isometric
 structures. Such  isogeometrical models of isofield interaction isotheories
 are in general nonlinear, nonlocal and
 nonhamiltonian and contain a very large class of local anisotropies
 and inhomogeneities induced  by four fundamental  isostructures:
 the partition of unity,  nonlinear isoconnection, distinguished
 isoconnections and isometric.

The novel geometric profile emerging from all the above studies is rather
 remarkable inasmuch as the first class of all isotopies herein
considered (called Kadeisvili's Class I
 [129,130]) preserves
the abstract axioms of conventional formulations, yet permits a clear
broadening of their applicability, and actually result to be
''directly universal'' 
 [223] for a number of possible well
 behaved nonlinear, nonlocal and nonhamiltonian systems. In turn, this
 permits a number of geometric unification such as that of all possible
 metrics (on isospaces with trivial nonlinear isoconnection structure)
 of a given dimension into Santilli's isoeuclidean metric, the
 unification of  exterior and interior gravitational problems despite
 their sizable structural differences and other unification.
\vskip20pt
\centerline{Chapter 12. {\bf Isobundle Spaces}}
\vskip10pt
This chapter serves the twofold purpose of establishing of abstract index
denotations and starting the geometric backgrounds of isotopic locally
an\-i\-sot\-rop\-ic extensions of the isoriemannian spaces which are used in
the next chapters of the work.

\vskip20pt
\centerline{Chapter 13. {\bf The Isogeometry of Tangent Iso\-bund\-les}}
\vskip10pt
The aim of this Chapter is to formulate some results in the isogeometry of
tangent isobundle, t--isobundle, $\widehat{TM}$ and to use them in
order to develop the geometry of Finsler and Lagrange isospaces.
\vskip20pt
\centerline{Chapter 14. {\bf Locally Anisotropic and
 Inhomogeneous Isogravity}}
\vskip10pt
The conventional Riemannian geometry can be generally assumed to be exactly
valid for the exterior gravitational problem in vacuum where  bodies can
 be well approximated as being massive points, thus implying the validity
 of conventional and calculus.

On the contrary, there have been serious doubt dating back to E. Cartan
 on the same exact validity of the Riemannian geometry for interior
gravitational problem because the latter imply internal effects which are
 arbitrary nonlinear in the velocities and other variables, nonlocal
 integral and of general non--(first)--order Lagrangian type.

Santilli 
 [221,222,224,223]
 constructed his isoriemannian geometry and proposed the related
 isogravitation theory precisely to resolve the latter shortcoming. In
 fact, the  isometric acquires an arbitrary functional; dependence thus
 being   able to   represent directly the locally anisotropic and
 inhomogeneous character of interior gravitational problems.

A remarkable aspect of the latter advances is that they were achieved by
 preserving the  abstract geometric axioms of the exterior gravitation.
 In fact, exterior and interior gravitation are unified in the above
geometric approach and are merely differentiated by the selected unit,
 the trivial value $I=diag(1,1,1,1)$ yielding the conventional gravitation
 in vacuum while more general realization of the unit yield interior
 conditions under the same abstract axioms (see ref.
 [129,130] for an independent study).

A number of applications of the isogeometries for interior problems have
already been  identified, such as (see ref. 
 [224] for an
 outline): the representation of the local variation of the speed of light
 within physical media such as atmospheres or chromospheres; the
 representation of the large difference between cosmological redshift
 between certain quasars and their associated galaxies when physically
 connected according to spectroscopic evidence; the initiation of the
 study of the origin  of the   gravitation via its identification with
 the field originating the mass of elementary  constituents.

As we have shown 
 [269,261,254,270]
the low energy limits of string
 and superstring theories  give  also rise
 to models of (super)field interactions
 with locally anisotropic and even higher order  anisotropic interactions.
The N--connection field can be treated as a corresponding nonlinear
 gauge field managing the dynamics of ''step by step'' splitting (reduction)
 of higher dimensional spaces to lower dimensional ones. Such (super)string
 induced (super)gravitational models have a generic local anisotropy
 and, in consequence, a more sophisticate  form of field equations and
 conservation laws and of corresponding theirs stochastic and quantum
 modifications. Perhaps similar considerations are in right for isotopic
 versions of sting theories. That it is why we are interested in a study
 of models of isogravity with nonvanishing nonlinear isoconnection,
 distinguished isotorsion and, in general, non--isometric fields.
\end{document}